\documentclass[aps,plb,letterpaper,twocolumn,reprint,preprintnumbers,
floatfix,nofootinbib,showpacs]{revtex4-1}
\usepackage{graphicx}
\usepackage{dcolumn}
\usepackage{bm}
\usepackage{epsfig,cancel}
\usepackage{epstopdf}
\usepackage{amsmath}
\usepackage{amssymb}
\usepackage{graphicx,bm}
\usepackage{color}
\usepackage{hyperref}
\usepackage{url}
\usepackage{breakurl}

\newcommand{\be}{\begin{equation}}
\newcommand{\ee}{\end{equation}}
\newcommand{\bea}{\begin{eqnarray}}
\newcommand{\eea}{\end{eqnarray}}

\def\mG{m_1}

\usepackage{color}

\begin{document}

\title{Kaluza-Klein graviton phenomenology for warped compactifications, and the 750 GeV diphoton excess}
\author{Steven B. Giddings}
\author{Hao Zhang}
\affiliation{Department of Physics, University of California, Santa Barbara, California 93106, USA}

\begin{abstract}
A generic prediction of scenarios with extra dimensions accessible in TeV-scale collisions is the existence of Kaluza-Klein excitations of the graviton.  For a broad class of strongly-warped scenarios one expects to initially find an isolated resonance, whose phenomenology in the simplest cases is described by a simplified model with two parameters, its mass, and a constant $\Lambda$ with units of mass parameterizing its coupling to the Standard Model stress tensor.  These parameters are in turn determined by the geometrical configuration of the warped compactification.  We explore the possibility that the 750 GeV excess recently seen in 13 TeV data at ATLAS and CMS could be such a warped Kaluza-Klein graviton, and find a best-fit value $\Lambda\approx 60$ TeV.  We find that while there is some tension between this  interpretation and data from 8 TeV and from the dilepton channel at 13 TeV, it is not strongly excluded.  However, in the simplest scenarios of this kind, such a signal should soon become apparent in both diphoton and dilepton channels.
\end{abstract}

\maketitle

\section{Introduction}
\label{sec:intro}
One fascinating possibility for physics at or near the TeV scale is that of 
gravity in extra dimensions.  Depending on the geometry of the extra 
dimensions, such a scenario can predict new relations\cite{ArkaniHamed:1998rs} 
between the fundamental Planck scale, where gravity becomes strongly coupled, and 
the weak scale, shedding new light on the hierarchy problem.  A generic 
feature of these scenarios is the presence of  Kaluza-Klein (KK) excitations 
of the graviton, which could be a first signature of such physics. 

An
excess in the diphoton invariant mass spectrum has recently been reported by both 
the ATLAS and CMS collaborations at the 13 TeV LHC \cite{ATLAS-CONF-2015-081,
CMS-PAS-EXO-15-004}, prompting a 
flood of $\approx 200$ 
papers
proposing its interpretation.\footnote{For an recent extensive list, see \cite{Aparicio:2016iwr}.}  However, most of these have focussed on the case of a 
spin-zero resonance (though \cite{Franceschini:2015kwy,Low:2015qep,Arun:2015ubr,
Han:2015cty,Kim:2015ksf,Martini:2016ahj,Geng:2016xin} consider aspects of spin-two resonances).  

In particular, it was argued in \cite{Franceschini:2015kwy} that a minimal KK graviton scenario is ruled out by the absence of a dilepton signal.\footnote{The {\sl Particle Data Book} \cite{Agashe:2014kda} also states a lower bound on KK gravitons of $\sim 2$ TeV, but this bound is strongly parameter dependent.}
However, given the significant implications and interest if the 750 GeV excess were the signal of extra 
dimensions, it is worth taking a careful look at this interpretation and bounds on it. 
This is one focus of this paper.  At the same time, we describe 
a simple and natural way to discuss the phenomenology of a KK graviton from 
a generic extra-dimensional scenario.  In general, such scenarios will be ``warped;" 
a simplest example of this phenomenon is exhibited in the Randall-Sundrum model\cite{Randall:1999ee}, 
but much more general possibilities exist in extra-dimensional theories such as 
string theory.  Indeed, if the 750 GeV excess is a KK graviton, the presence of an isolated resonance at this mass scale 
would indicate a strongly warped scenario, as opposed to one with purely large extra dimensions\cite{ArkaniHamed:1998rs,Antoniadis:1998ig}.  
We describe a simple and general parameterization 
of general warped compactifications, and of the phenomenology of their KK graviton 
modes.  In particular, the low-energy phenomenology of the simplest models -- with a warped metric and SM matter concentrated at a single point in the extra dimensions -- 
is naturally described 
in terms of the mass of the lightest KK graviton, and a single parameter $\Lambda$ with mass 
dimension one, which characterizes the minimal coupling of this KK graviton to the 
stress tensor of Standard Model (SM) matter.  

This paper will explore the interpretation of the LHC13 data as fixing the value 
of this coupling parameter, and will examine other constraints on it.  We find 
that the combined  13 TeV excess gives $\Lambda\approx\,60$ TeV, and while 
other constraints begin to have tension with this, such a coupling strength is 
not strongly ruled out.  Also, in the simple example of the Randall-Sundrum model, the 
relevant parameters are apparently not ruled out by the absence of a radion signal, since the radion is 
expected to be light ($<4$ GeV) and very weakly coupled.
In turn, in the simplest warped compactification models, the coupling $\Lambda$ determines the value 
of the higher-dimensional Planck mass.  We also find that, again in the simplest 
models, such a value for $\Lambda$ should be confirmed, or ruled out, in both 
the upcoming diphoton data and in that for dileptons.

\section{Warped compactifications and TeV(ish)-scale gravity}

\subsection{General warped compactifications}

Generic extra-dimensional configurations, {\it e.g.} in string theory, 
are warped, due to the possible presence of frozen-in fluxes and branes.  This 
means that the $D$-dimensional spacetime metric takes the form
\be \label{warpmet}
ds^2= e^{2A(y)} dx_4^2 + g_{ab}(y) dy^a dy^b\ ,
\ee
where $A$ is a function of the  $n=D-4$ compact coordinates $y^a$, 
$dx_4^2$ is our four-dimensional, nearly-Minkowski metric, and $g_{ab}$ 
are the compact components of the metric.  
Such compactifications can lead to interesting new explanations for the 
relative sizes of the fundamental Planck scale, the four-dimensional Newton's 
constant, and the weak scale.  In particular,  if $M_D$ is the $D$-dimensional 
Planck scale, then the four-dimensional Planck scale, $M_4\simeq 2.4\times 
10^{18}$ GeV, related to Newton's constant by $M_4^2=1/(8\pi G_N)$, is given 
by\footnote{Herein we mainly use the conventions of the large extra dimensions 
section of \cite{Agashe:2014kda}, which differ from those of the warped extra dimensions 
section; for related broad discussion of warped 
compactification parameters see \cite{Giddings:2008gr}.}
\be\label{planckrel}
\frac{M_4^2}{M_D^2} = V_W \left(\frac{M_D}{2\pi}\right)^{D-4} 
\ee
where
\be
{V_W= \int d^ny e^{2A(y)} \sqrt{g(y)}}\ 
\ee
is called the {\it warped volume}.  From \eqref{planckrel} one sees 
that the fundamental Planck scale $M_D$ can be in the $TeV$ range -- so with large ratio $M_4/M_D$ --
either due to large warp factor $e^A$, or to large volume $\int d^n y \sqrt g$, or to some 
combination of the two.  This recasts the hierarchy problem as one 
of explaining the large warped volume.  

Limiting cases are the large-extra dimensions scenarios of Arkani-Hamed, Dimopoulos, and Dvali\cite{ArkaniHamed:1998rs,Antoniadis:1998ig}, and the toy warped  model of Randall and Sundrum\cite{Randall:1999ee}, 
but a continuum of possibilities exist.  In the simplest scenarios, SM 
matter is taken to reside on a $3+1$ dimensional subspace of the full 
geometry, commonly defined by a brane such as the D-branes of string 
theory, although extra-dimensional structure for gauge fields and matter 
is also possible.  

If $M_D$ is in a range near the $TeV$ scale, a variety of important 
new prospects for phenomenology present themselves; one can in 
particular  find new states below $M_D$ arising as higher-dimensional 
Kaluza-Klein modes.   For example in the large-radius
\cite{ArkaniHamed:1998rs,Antoniadis:1998ig} limiting case, with vanishing warping $A$ 
and a toroidal compact manifold, these KK modes have masses 
$m_{KK}\sim 1/R_i$ where $R_i$ are the radii of the torus.  More 
generally the KK masses are determined by  characteristic  
radii of curvature of the geometry, so may depend  on the geometry of $g_{ab}$ 
and/or on the variation scale of the warp factor $A(y)$.  This means 
it is useful to parameterize such phenomenology simply by the lightest 
KK graviton mass, $\mG$, which can have complicated dependence 
on the internal geometry.  There can be other modes in a similar mass 
range, such as four-dimensional scalars arising from low-energy modes 
of $g_{ab}$ in the extra dimensions; these moduli modes include the 
``radion," arising from an overall rescaling of the internal metric.

For the time being we focus on phenomenology of the lightest KK graviton.  
This state can be described by expanding the four-dimensional part of the metric as
\be
{g_{\mu\nu}(x,y)} = e^{2A(y)} \left[\eta_{\mu\nu} + \kappa h_{\mu\nu}(x,y)\right]
\ee
where $\kappa^2=(2\pi)^{D-4}/M_D^{D-2}$ gives the gravitational coupling, 
and the expansion of the metric perturbation in KK modes is
\be
h_{\mu\nu}(x,y)=\sum_{N=0}^\infty h_{N\mu\nu}(x) \phi_N(y)\ .
\ee
$\phi_N(y)$ are the internal wavefunctions of the KK modes, and $N=0$, 
with constant wavefunction, gives the 4d graviton.  In the simplest scenarios, 
focussing on the graviton yields significant predictivity, since in these scenarios gravitons universally 
couple to the stress tensor $T_{\mu\nu}$ of the SM.  In particular, consider 
a higher-dimensional lagrangian including the Einstein-Hilbert term and 
SM matter localized on the brane,
\be
S=\frac{M_D^{D-2}}{(2\pi)^{D-4}} \int d^DX \sqrt{-g}\frac{\cal R}{2} + 
\int d^DX \sqrt{-g}\,  \delta^n(y){\cal L}_{\rm SM} +\cdots
\ee
with $\delta^n(y)$ localizing  to $y^a=0$. {We choose the scale
of $x^\mu$ to set $A(0)=0$.} If one chooses to normalize 
$\phi_1$ so $h_{1\mu\nu}$ has a 4d canonical kinetic term,  one finds 
a four-dimensional interaction lagrangian 
\be\label{intlag}
{\cal L}_1= -{1\over \Lambda} h_{1\mu\nu}(x) T^{\mu\nu}(x)\ .
\ee
Here the dimension-one constant $\Lambda$ is given by the gravitational 
coupling $\kappa$ and the wavefunction $\phi_1$ at the Standard Model 
location $y^a=0$,
\be \label{Lambdarel}
\Lambda = \frac{2}{\kappa\phi_1(0)} = M_D \left(\frac{M_D r_1}{2\pi}\right)^{n/2}\ .
\ee
In the last equality  a simple parameterization of the extra-dimensional wavefunction is 
introduced as a radius, $\phi_1(0)= 1/r_1^{n/2}$; this radius $r_1$ 
characterizes the typical density of the KK wavefunction near
the brane.

The bottom line is simple:  the couplings of the lowest KK graviton to SM 
fields may be parameterized by a single mass scale $\Lambda$, and in the simplest scenarios this state's 
phenomenology is largely determined by this parameter and its mass 
$\mG$.  We treat these as free parameters, though return to comment 
on their possible sizes later; one can think of this as defining a ``simplified model" for KK graviton phenomenology.  

\subsection{The Randall-Sundrum model}

The Randall-Sundrum two-brane model \cite{Randall:1999ee} RS1 provides 
an illustrative toy model of the preceding discussion.  It may be described by 
a five-dimensional metric (compare \eqref{warpmet})
\be
ds^2= e^{2ky} dx_4^2 + dy^2\ 
\ee
where $y$ ranges from $0$ (``IR" or ``SM brane") to $\pi R$ (``UV brane").
This differs from the original parameterization\cite{Randall:1999ee} in terms 
of coordinates $x',\phi$ by 
\be
x' = e^{\pi k R} x\quad ,\quad \phi=\pi  -y/R\ ;
\ee
in \cite{Randall:1999ee} $R$ was called $r_c$.
Then, the four- and five-dimensional Planck masses are related by (compare 
\eqref{planckrel})
\be\label{rsm4m5}
M_4^2=\frac{M_5^3}{2\pi} \int_0^{\pi R} dy e^{2ky} = 
\frac{M_5^3}{4\pi k}\left(e^{2\pi k R} -1\right)\ .
\ee
The mass of the lowest graviton KK mode is \cite{Davoudiasl:1999jd}
\be\label{massrs}
\mG= x_1 k
\ee
where $x_1=3.83$ is the first zero of the Bessel function $J_1$.  
The parameter $k$ also determines the falloff radius $r_1\propto 1/k$, 
so we find\cite{Davoudiasl:1999jd} (compare \eqref{Lambdarel}) 
\be\label{rsl_m5}
\Lambda = \frac{M_5}{2}\sqrt{\frac{M_5}{4\pi k}\left(1-e^{-2\pi k R}\right)} 
\simeq M_5 \sqrt{\frac{M_5}{16 \pi k}}\ 
\ee
since typically $kR\gg1$.
In this RS1 context, $\Lambda$ has instead been called $\Lambda_\pi$, 
and is related to  ${\overline M}_{{Pl}}=M_4/2$ of \cite{Davoudiasl:1999jd} by 
\be
\Lambda=\Lambda_\pi=e^{-\pi k R}{\overline M}_{{Pl}}\ .
\ee

\section{The signal}
\label{sec:sig}
We now turn to a discussion of the intriguing possibility that a warped KK 
graviton $G_1^*$ could be the source of the 750 GeV diphoton excess, and to other constraints 
on such a scenario.  In this paper we focus on the simplest case where the 
SM is restricted to a brane at $y^m=0$, as described above, although more 
general scenarios can be explored with Standard Model fields extending 
into the extra dimensions.

If a KK graviton is the source of the 750 GeV excess, this implies that the warping 
is significant, since in a pure large extra dimensions scenario\cite{ArkaniHamed:1998rs,Antoniadis:1998ig} there would be a 
near-continuum of KK excitations on these energy scales. 
As we have discussed, $\mG=750$ GeV is then related to a characteristic 
curvature scale of the extra dimensions; for the example of RS1 this would 
imply $k=196$ GeV.

The remaining free parameter is $\Lambda$ in \eqref{intlag}, which can be 
fixed by matching to the signal cross section $\sigma\left(pp\to
G^*_1\to \gamma\gamma\right)$. A 1.755 $K$-factor for 13 TeV, from the 
next-to-leading order (NLO) QCD corrections\cite{Mathews:2005bw,Li:2006yv,
Kumar:2009nn,Gao:2010bb}, is used in our work. We 
calculate the leading order (LO) graviton production cross section 
with MadGraph5 \cite{Alwall:2014hca} with CT14llo parton distribution function 
(PDF) \cite{Dulat:2015mca}. The renormalization scale ($\mu_R$) and the 
factorization scale ($\mu_F$) are fixed to be $\mu_R=\mu_F=750$ GeV. 
Since we use the NLO QCD $K$-factor in calculating the inclusive 
cross section, the (renormalization and factorization) scale 
dependence uncertainty is estimated to be suppressed to 9.3\% \cite{Li:2006yv}. The 
uncertainties from the choice of the PDFs are estimated by also performing calculations
 with CTEQ6L1\cite{Pumplin:2002vw} 
and MSTW2008LO PDFs\cite{Martin:2009iq}. 
We then find, including the combined uncertainties,
\be
\sigma\left(pp\to G^*_1\right)=7.74^{+1.43}_{-1.10}{\text{pb}}\times
\left(\frac{10{\text{TeV}}}{\Lambda}\right)^2.
\ee
Here we estimate the central value by taking the algebraic average of the central values from the different PDFs, and 
the error region is the region covered by the scale uncertainties with different PDFs.

\begin{figure}[!htb]
\includegraphics[scale=0.4,clip]{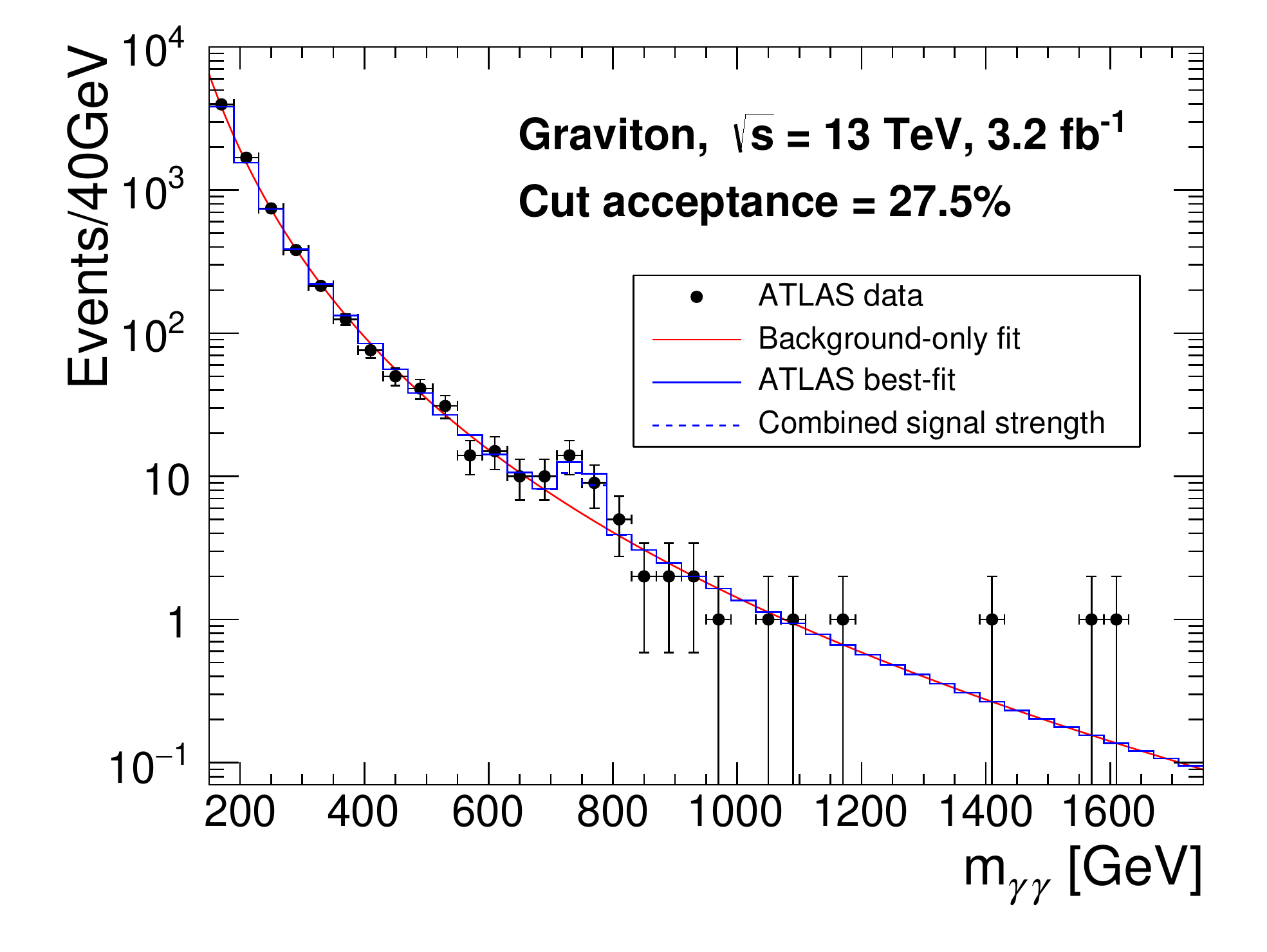}
\caption{The best-fit result to the 13 TeV LHC diphoton excess
with a lightest KK graviton. The narrow 
width approximation is used in the fit.
The solid blue line is our best-fit  to the 
13 TeV ATLAS data. The dashed blue line 
is the best-fit  from a combination 
of the 13 TeV CMS results (from the  CMS
collaboration \cite{CMS-PAS-EXO-15-004})
and our best-fit of the 13 TeV ATLAS data.
\label{fig:fit} }
\end{figure}

To fit the excess from the ATLAS collaboration, we generate
parton level events using MadGraph5 \cite{Alwall:2014hca} 
with CT14llo parton distribution function (PDF) \cite{Dulat:2015mca}. 
$pp\to G_1^*+{\text{n}} j$ events are generated to n=1. The MLM 
matching scheme is used to avoid  double counting in the parton 
showering \cite{Mangano:2006rw}. All parton level events are 
showered using PYTHIA6.4 with Tune Z2 parameter assignment 
\cite{Sjostrand:2006za,Field:2011iq}.
We use DELPHES3 to mimic  detector effects
\cite{deFavereau:2013fsa,Cacciari:2011ma}. More details 
of this fitting procedure can be found in \cite{Gao:2015igz,Zhang:2016pip}.
The best-fit result is shown in FIG. \ref{fig:fit}.
The best-fit unfolded cross section from ATLAS is $\sigma\left(pp\to
G^*_1\to \gamma\gamma\right)=13.7_{-5.1}^{+5.9}$ fb. 
The cut acceptance from our simulation is 27.5\% for 
a narrow width KK graviton.

\begin{figure}[!htb]
\includegraphics[scale=0.4,clip]{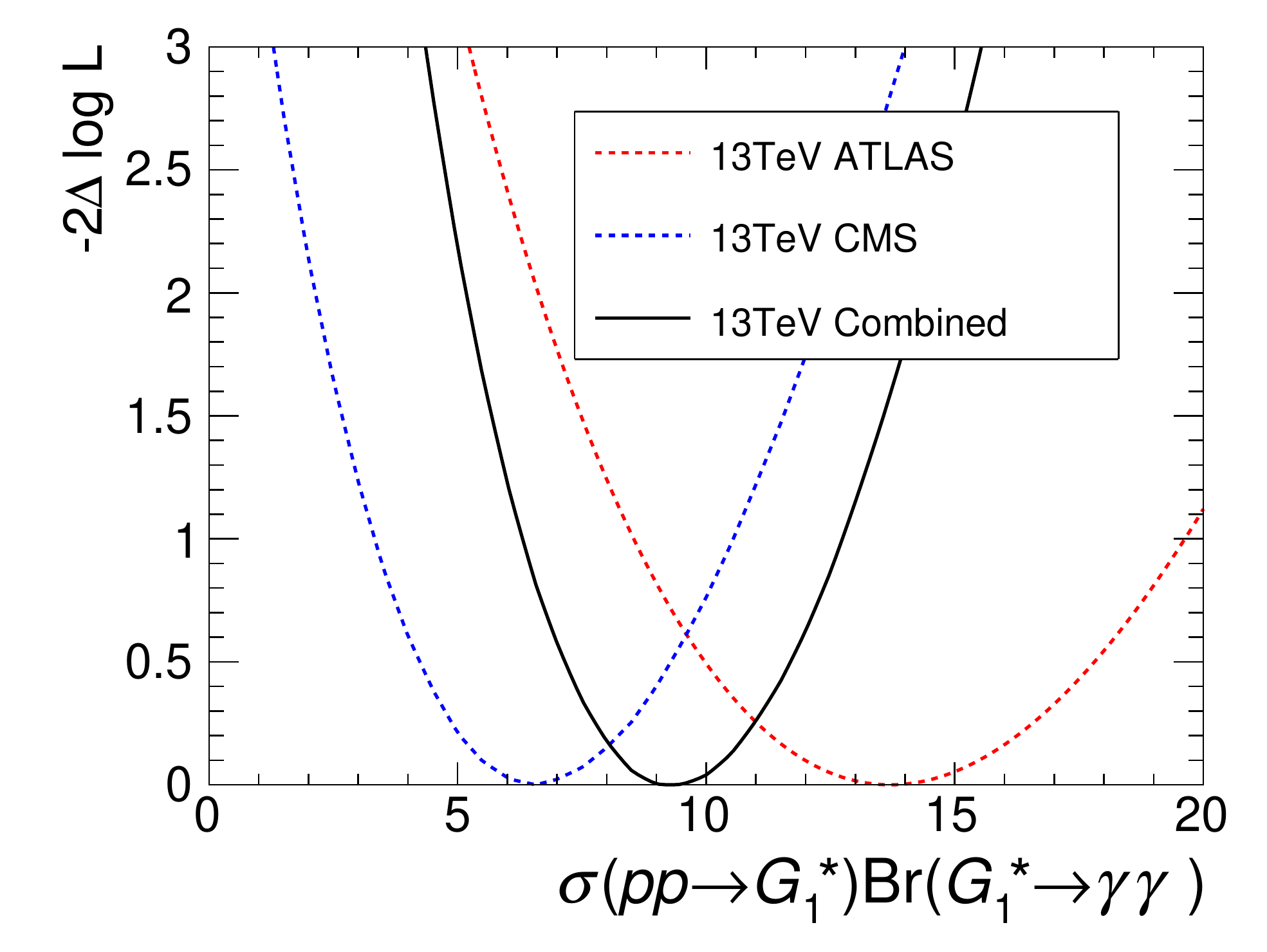}
\caption{The likelihood functions for the diphoton signal search. The
likelihood function of the ATLAS excess is from our fit. The 
likelihood function of the CMS excess
is extracted from the result presented in \cite{CMS-PAS-EXO-15-004}. 
We combine these results to give the result with the black solid line. 
The best-fit cross section of the 13 TeV LHC is $\sigma\left(pp\to
G^*_1\to \gamma\gamma\right)=9.3_{-2.9}^{+3.3}$ fb.
\label{fig:likelihood} }
\end{figure}

A best-fit result 
to the 13 TeV CMS data ($\sigma=6.6_{-3.3}^{+3.9}$ fb) was performed 
by the CMS collaboration.
The best-fit result 
after combining the 13 TeV ATLAS and 13 TeV CMS
results is $\sigma\left(pp\to
G^*_1\to \gamma\gamma\right)=9.3_{-2.9}^{+3.3}$ fb.  
The likelihood functions are shown in FIG. \ref{fig:likelihood}. 

The partial decay widths of the KK graviton to 
SM particles via the Lagrangian (\ref{intlag}) are given in \cite{Han:1998sg}. 
The partial decay width of the KK graviton to massless gauge bosons ($\Gamma_{V_0V_0}$),
massive gauge bosons ($\Gamma_{VV}$),
fermions ($\Gamma_{ff}$) and the Higgs boson ($\Gamma_{hh}$) are
\bea
\Gamma_{V_0V_0}&=&\frac{N_C\mG^3}{80\pi\Lambda^2}\\
\Gamma_{VV}&=&\delta\frac{\mG^3}{40\pi\Lambda^2}
\left(1-\frac{4m_V^2}{\mG^2}\right)^{1/2}\nonumber\\&&\times
\left(\frac{13}{12}+\frac{14m_V^2}{39\mG^2}+\frac{4m_V^4}{13\mG^4}\right),\\
\Gamma_{ff}&=&\delta\frac{N_C\mG^3}{160\pi\Lambda^2}
\left(1-\frac{4m_f^2}{\mG^2}\right)^{3/2}
\left(1+\frac{8m_f^2}{3\mG^2}\right),\\
\Gamma_{hh}&=&\frac{\mG^3}{480\pi\Lambda^2}
\left(1-\frac{4m_h^2}{\mG^2}\right)^{5/2},
\eea
where $N_C$ is a color factor which is 8 for gluons, 3 for 
quarks and 1 for colorless particles, and $\delta$ is $1/2$ for 
self-conjugate particles and 1 for other particles. The 
total width of a 750 GeV KK graviton is then
\be
\Gamma=0.39{\text{GeV}}\times\left(\frac{10{\text{TeV}}}{\Lambda}\right)^2 \ ,
\ee
justifying the narrow-width approximation.

The decay branching ratio to the diphoton final state is (see FIG. \ref{fig:rsbr})
\be
{\text{Br}}\left(G^*_1\to\gamma\gamma\right)=4.3\%,
\ee
As a result, the best-fit coupling scale from the ATLAS diphoton excess is 
\be
\Lambda_{\text{ATLAS}}\simeq50^{+12}_{-10}{\text{TeV}} 
\ee
and the CMS best-fit coupling scale is 
\be
\Lambda_{\text{CMS}}\simeq71_{-18}^{+22}{\text{TeV}}.
\ee
If we consider the combined results,  
the best-fit coupling scale  is 
\be
\Lambda_{\text{combined}}\simeq60_{-10}^{+12}{\text{TeV}}.
\ee
The PDF uncertainties and statistical uncertainty are 
combined independently in these error estimates.

If a warped KK graviton couples with this strength, one also 
expects signals in other channels\cite{Agashe:2014kda,
Franceschini:2015kwy}.  Therefore, we next turn to an examination 
of constraints from other LHC data.

\section{Constraints from  LHC Run-I}
\label{sec:run1}
\begin{figure}[!htb]
\includegraphics[scale=0.5,clip]{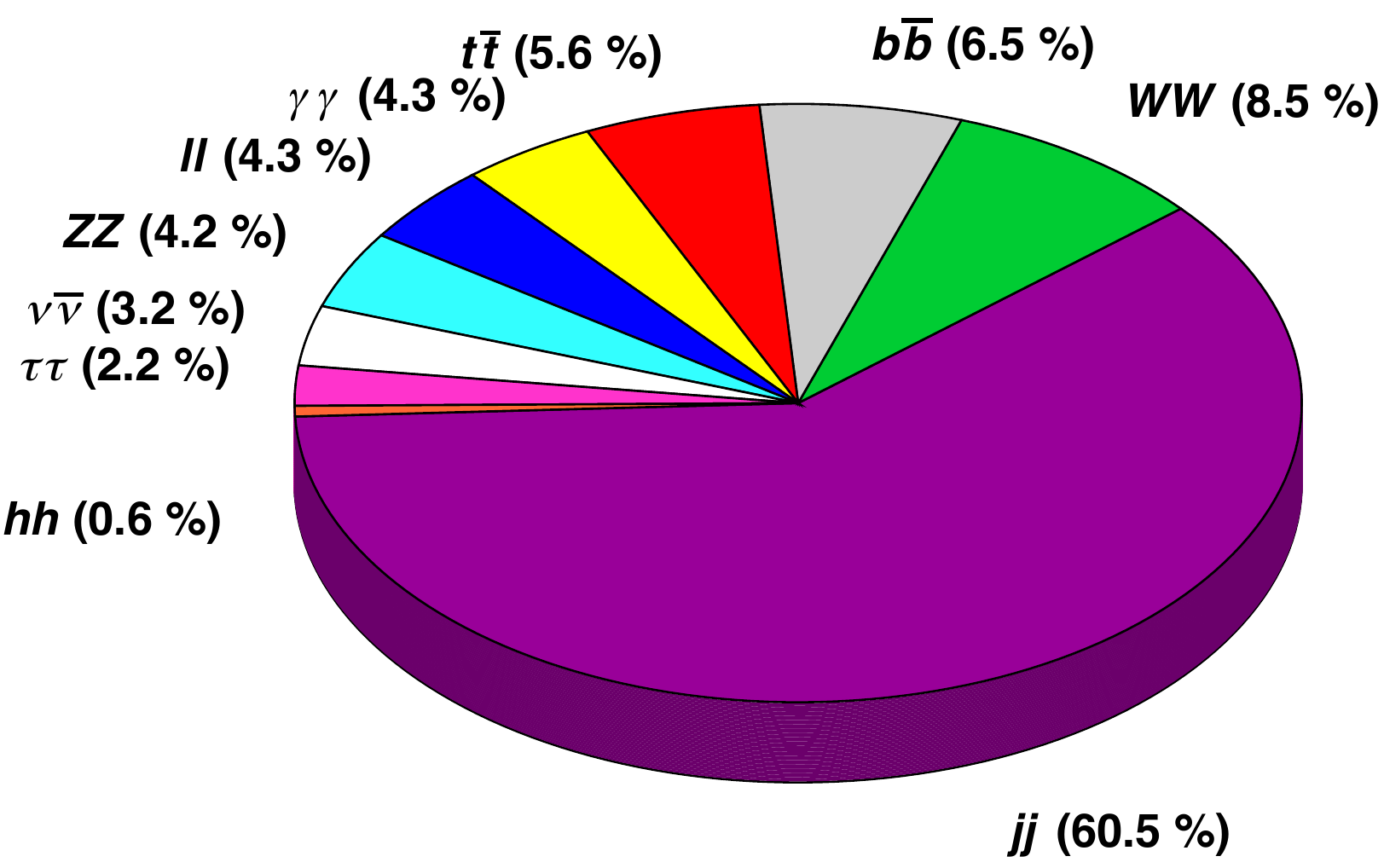}
\caption{The decay branching ratios of the 750 GeV KK graviton.
\label{fig:rsbr} }
\end{figure}

In this section, we examine constraints
on a warped KK graviton from direct searches performed on LHC Run-I data. 
From the decay branching ratios of a 750 GeV KK graviton shown in 
FIG. \ref{fig:rsbr}, we find that the most 
important decay channel is the dijet channel where here the jets are 
associated with gluons and up, down, charm and strange quarks. 
The constraint from this channel is weak due to the huge SM background. 
The 8 TeV 
inclusive cross section of the 750 GeV KK graviton is calculated 
with MadGraph5 with the parameters and PDFs described in the last section. 
The result is (with the $K$-factor 1.922 \cite{Gao:2010bb}, for 8 TeV)
\be\label{sigmatolambda}
\sigma\left(pp\to G^*_1\right)=2.01^{+0.47}_{-0.37}{\text{pb}}\times
\left(\frac{10{\text{TeV}}}{\Lambda}\right)^2.
\ee
In TABLE \ref{table:8tev}, we list constraints from  ATLAS and 
CMS at the 8 TeV LHC. The strongest constraint on $\Lambda$ comes from the 
8 TeV diphoton search by the CMS collaboration,
which rule out the parameter range $\Lambda<69$ TeV at 
95\% C.L. Thus the best-fit point to the 13 TeV ATLAS
diphoton data is excluded by this result. However, the 
CMS best-fit result and the combined fit result is still  
allowed within 1$\sigma$. 
\begin{table}[!htb]
\caption{The constraints on $\sigma\left(pp\to G^*_1\right)$ 
at 8 TeV LHC (95\% C.L. upper bound). The $\mathcal{A}$ in 
the table is the cut acceptance of the process.  The bounds for $\Lambda$ shown in the last column are calculated using the central value of eq.~\eqref{sigmatolambda}. }
\begin{center}
\begin{tabular}{c|c|c|c}
\hline\hline
&$\sigma\left(pp\to G^*_1\right)$Br~(pb)&$\sigma\left(pp\to G^*_1\right)$~(pb)&$\Lambda$~(TeV)\\
\hline
$jj$&$1.25/\mathcal{A}$ \cite{Aad:2014aqa}&$1.9/\mathcal{A}$&$10\mathcal{A}$\\
$VV$&$0.065$ \cite{Aad:2015ipg} $0.055$ \cite{CMS-PAS-EXO-12-022}&$1.2$&$13$\\
$t\bar t$&$0.52$ \cite{Aad:2015fna}, $0.4$ \cite{Khachatryan:2015sma} 
&$7.8$&$5.1$\\
$\gamma\gamma$&$0.0024$ \cite{Aad:2015mna}, $0.0018$ \cite{CMS-PAS-EXO-12-045}
&$0.042$&$69$\\
$\ell^+\ell^-$&$0.0011$ \cite{Aad:2014cka}, $0.0014$ \cite{Khachatryan:2014fba}
&$0.051$&$63$\\
$\tau^+\tau^-$&$0.01$\footnote{This is the constraint on a  spin-1 particle.
The spin-2 KK graviton will have a little larger cut acceptance and the exact constraint will 
be a little smaller but roughly the same.} \cite{Aad:2015osa}&$0.47$&$21$\\
$hh$&$0.045$ \cite{Aad:2015xja}, $0.041$ \cite{Khachatryan:2015yea}&$7.7$&$5.1$\\
\hline\hline
\end{tabular}
\end{center}
\label{table:8tev}
\end{table}

We can also consider the signal and bounds from ATLAS 
and CMS collaborations separately. The 8 TeV bound from 
ATLAS collaboration is $\Lambda>63$ TeV from the dilepton 
channel. So the ATLAS signal is excluded by its 8 TeV result.
The 8 TeV bound from the CMS collaboration is $\Lambda>69$ TeV
from the diphoton channel, so  the best-fit result of the CMS signal 
is just at the exclusion bound. Again, if we consider the uncertainties 
from the PDFs, as shown in TABLE \ref{table:8tevpdf}, it is 
seen that the CMS best-fit and the combined result are certainly 
allowed within 1$\sigma$ when these uncertainties are included. 
The 1$\sigma$ ATLAS best-fit region is at the edge of the 
exclusive bound from the ATLAS 8 TeV results when we considered 
these PDF uncertainties. 
\begin{table}[!htb]
\caption{The constraints on $\Lambda$ 
at 8 TeV LHC (95\% C.L. upper bound), with  PDF uncertainties.}
\begin{center}
\begin{tabular}{c|ccccccc}
\hline\hline
&$jj$&$VV$&$t\bar t$&$\gamma\gamma$&$\ell^+\ell^-$&$\tau^+\tau^-$&$hh$\\
\hline
$\Lambda$ (TeV)&$10^{+1.2}_{0.9}\mathcal{A}$&$13^{+1.5}_{-1.2}$&
$5.1^{+0.6}_{-0.5}$&$69^{+8}_{-6}$&$63^{+7}_{-6}$&$21^{+2}_{-2}$&$5.1^{+0.6}_{-0.5}$\\
\hline\hline
\end{tabular}
\end{center}
\label{table:8tevpdf}
\end{table}

\section{Constraints from  LHC Run-II}
\label{sec:run2}
One should also examine the dilepton constraints  from the 13 TeV LHC on a warped KK graviton.
Both ATLAS and CMS collaborations search for an exotic spin-1 resonance in
the dilepton final state \cite{ATLAS-CONF-2015-070,CMS-PAS-EXO-15-005}.
The upper bounds are summarized in TABLE \ref{table:13tev2l}.
\begin{table}[!htb]
\caption{The upper bounds on $\sigma\left(pp\to Z^\prime\right){\text{Br}}
\left(Z^\prime\to\ell\ell\right)$ from the 13 TeV LHC. The unit in this table is fb.}
\begin{center}
\begin{tabular}{c|ccc}
\hline\hline
Channel &~~ ~~$ee$~~ & ~~$\mu\mu$~~ & ~~combined~~\\
\hline
~~13 TeV ATLAS \cite{ATLAS-CONF-2015-070}~~ & ~~~~$6.2$~~ & ~~$13.8$~~ 
& ~~$5.6$~~ \\
~~13 TeV CMS \cite{CMS-PAS-EXO-15-005}~~~~~~~& ~~~~$3.5$~~ & ~~
$8.5$~~ & ~~$2.9$~~ \\
\hline\hline
\end{tabular}
\end{center}
\label{table:13tev2l}
\end{table}To give a constraint for a graviton, we generate the 
signal events by the same method as in Sec. \ref{sec:sig} but require the graviton to
decay into dileptons.  We also simulate a 750 GeV $Z^\prime$ for comparison, and infer the graviton bounds using the ratio between the cut acceptances, which should suppress uncertainties.
 The ratios between
 graviton and  $Z^\prime$ cut acceptances ($\epsilon_{X}^{\ell\ell}$) 
from our simulations are
\bea
&&{\text{ATLAS:}}~~R_{\mu\mu}=\frac{\epsilon_{G^*_1}^{\mu\mu}}{\epsilon_{Z^\prime}^{\mu\mu}}
=1.12,~R_{ee}=\frac{\epsilon_{G^*_1}^{ee}}{\epsilon_{Z^\prime}^{ee}}
=1.35,\\
&&{\text{CMS:}}~~R_{\mu\mu}=\frac{\epsilon_{G^*_1}^{\mu\mu}}{\epsilon_{Z^\prime}^{\mu\mu}}
=1.16,~R_{ee}=\frac{\epsilon_{G^*_1}^{ee}}{\epsilon_{Z^\prime}^{ee}}
=1.23.
\eea
The cut acceptances for the graviton are larger than for the $Z^\prime$ due to 
the final state leptons being more central in the detector on average, which 
was also noted for the  8 TeV LHC (e.g., see Ref \cite{Khachatryan:2014fba}).
The strongest constraints are from the $ee$-channel. We  show
the constraints on a warped KK graviton from such 13 TeV dilepton searches in 
TABLE \ref{table:13tev2lg}. We only show the result with CT14llo PDF
since the constraints here are also from the 13 TeV LHC.
\begin{table}[!htb]
\caption{The bounds on $\sigma\left(pp\to G_1^*\right)
{\text{Br}}\left(G^*_1\to\gamma\gamma\right)$ 
and $\Lambda$ at 13 TeV LHC.}
\begin{center}
\begin{tabular}{c|cccc}
\hline\hline
Channel&ATLAS ($ee$)&ATLAS ($\mu\mu$)&CMS ($ee$)&CMS ($\mu\mu$)\\
\hline
$\sigma$ (fb)&$9.2$&$24.6$&$5.7$&$14.7$\\
$\Lambda$ (TeV)&$60$&$37$&$76$&$48$\\
\hline\hline
\end{tabular}
\end{center}
\label{table:13tev2lg}
\end{table}
We see from the result that the combined best-fit result 
from the 13 TeV ATLAS and 13 TeV CMS diphoton
data is nearly excluded by the 13 TeV dilepton constraints (CMS $ee$
channel), at 98\% C.L. The best-fit 
result from the ATLAS signal is  excluded by the 13 TeV CMS 
dielectron channel with 98.8\% C.L., and excluded by the 13 TeV ATLAS 
dielectron channel with 96.8\% C.L. The best-fit result from the 13 TeV CMS diphoton
data is however (with less than 1$\sigma$ uncertainty) still consistent with the 13 TeV 
dilepton constraints.

The CMS report\cite{CMS-PAS-EXO-15-004}  also combines the
8 TeV results with the 13 TeV results. The best-fit  
to the combined 8 TeV and 13 TeV CMS diphoton data is 
\bea
&\sigma&\left(pp\to
G^*_1\to \gamma\gamma\right)=4.5_{-1.7}^{+1.9} ~{\text{fb}},\\
&\Lambda&=86_{-17}^{+20} ~{\text{TeV}},
\eea
which is  consistent with the constraint from the 13 TeV LHC 
dilepton data. 

We also conclude from this discussion that if  the 750 GeV excess 
is confirmed by the next round of data, and its source is a KK graviton 
in the simple type of warped compactification that we have considered,
the resonance should also be found very soon in the dilepton 
channel.\footnote{Alternately, non-trivial bulk structure of the SM 
fermions in a more complicated scenario could alter this result; 
see the discussion below.}

\section{Distinguishing the graviton from a spin-0 resonance}
It is well known that the final state diphoton angular distribution 
can be used to investigate the spin of a resonance.
\begin{figure}[!htb]
\includegraphics[scale=0.4,clip]{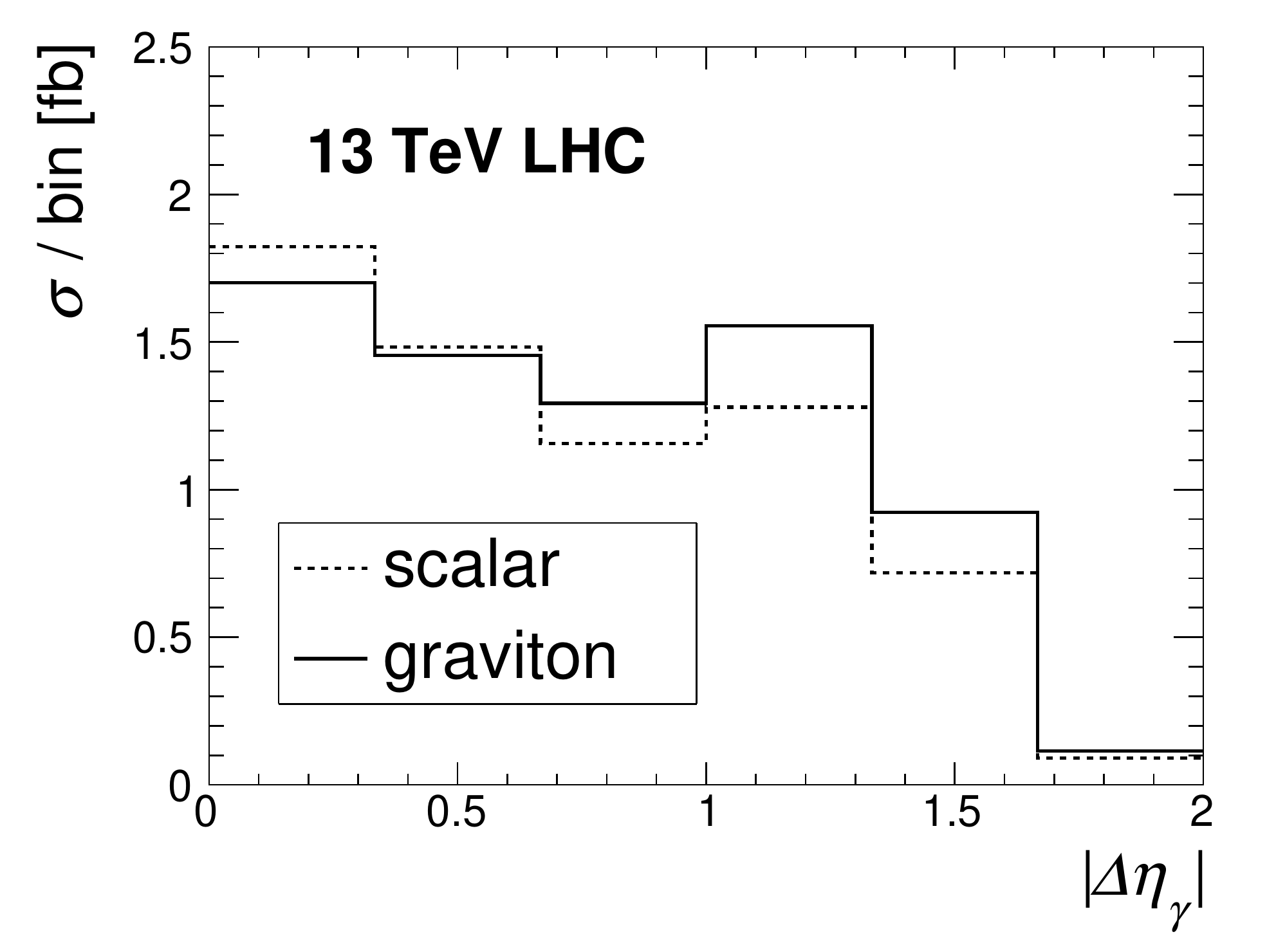}
\caption{The best-fit results for the LHC Run-II diphoton excess
with the lightest KK graviton. We use the narrow 
width approximation. 
\label{fig:0or2} }
\end{figure}In this 
section, we estimate the  luminosity which is needed
to measure the spin of this diphoton resonance. For an (optimistic) 
example we assume the 
signal strength is the best-fit value to the ATLAS excess. In this 
preliminary estimate, we only generate the $pp\to \gamma\gamma$
SM background events with at most one additional jet with MadGraph5, 
PYTHIA6.4 and MLM matching scheme. The background
strength is rescaled to the ATLAS background. In addition to the 
cuts used by the ATLAS collaboration \cite{ATLAS-CONF-2015-081},
we require that the diphoton invariant mass satisfy
\be
|m_{\gamma\gamma}-750~{\text{GeV}}|<40~{\text{GeV}}.
\ee

The distribution of the difference between the pseudo-rapidity of the leading 
and subleading photons in each event is shown in FIG. \ref{fig:0or2}.
To distinguish a spin-2 resonance from a spin-0 resonance
with 99\% C.L., roughly 50 fb$^{-1}$ integrated luminosity is needed 
at 13 TeV. Of course, assuming the excess persists, more data will 
be needed to distinguish between spin 0 and spin 2 if the estimated 
signal cross section decreases. Note that the CMS collaboration separates their events into the 
``EBEB'' (containing photon pairs where both candidates are reconstructed 
in the electromagnetic calorimeter (ECAL) barrel) and ``EBEE''
(containing photon pairs where one of the candidates is reconstructed 
in the ECAL endcaps) categories. Their data shows that there is 
a significant (in fact, more) contribution to the signal from the EBEE
category. This could  hint that a spin-2 resonance is favored
by the excess, but more data is needed to draw a conclusion.

\section{Relating phenomenological to fundamental parameters}

As we have described, measurement of the mass of the lightest KK graviton 
can be thought of as determining a typical curvature radius scale of the extra-dimensional 
geometry.  As a specific example, eq.~\eqref{massrs} shows that a mass of 
750 GeV fixes $k=196~{\text{GeV}}$ in RS1.

The strength of the signal, parameterized by $\Lambda$, is a combination 
of the higher-dimensional Planck mass, and the scale $r_1$ determining 
the density of the wavefunction in the extra dimensions, as in \eqref{Lambdarel}. 
If $r_1$ is the same scale as $1/\mG$, as might be typically expected, this 
then determines $M_D$.  Specifically, in the example of RS1, this happens 
through \eqref{rsl_m5}, which with $\Lambda=60$ TeV, determines $M_5=33$ TeV.  

Finally, the warped volume is then determined by the ratio $M_4/M_D$, as 
in \eqref{planckrel}.  In the RS1 example this becomes \eqref{rsm4m5}, 
which determines $kR\simeq {9.7}$.

\section{Moduli and Radion}
In general warped compactifications, there will also be light scalar fields 
arising from deformations of the compact metric $g_{ab}$ which correspond 
to moduli.  In general, the dynamics of the extra dimensions must provide a 
potential that gives mass to these deformations, for realistic phenomenology.  
We won't consider the full story of such moduli, which can be complicated, here,
 but instead briefly illustrate such considerations in the toy model of RS1.  

In RS1 there is a single light scalar radion arising from the $yy$ component of the metric.  
A simple stabilization mechanism was introduced  by Goldberger and Wise in 
\cite{Goldberger:1999uk}. This gave a radion mass which is determined in terms 
of $k$ and $R$, as well as certain dimension 3/2 vacuum expectation value
(vev) parameters $v_h$, $v_v$, 
as\cite{Goldberger:1999un}
\be
m_{rad} = \frac{v_v}{\sqrt 3 \pi M_5^{3/2}} \frac{1}{R} \log\left(\frac{v_h}{v_v}\right)\ 
\ee
(here $v_v, v_h$ are rescaled compared to \cite{Goldberger:1999un});
corrections including back reaction are given in \cite{Csaki:2000zn}. 
Since $kR\sim 10$ to generate the hierarchy, a subplanckian vev scale\cite{Goldberger:1999uk} $v_v/M_5^{3/2}<1$ implies a 
radion mass in Goldberger/Wise-stabilized RS1 that is well below that of the 
KK graviton; with the parameters inferred above, we find $m_{rad}\lesssim 4$ GeV.
Searches for such a radion provide another test for such warped compactifications, though it is difficult to make precise and general statements for the general such compactification, where moduli phenomenology depends on the details of the stabilization mechanism.

To illustrate the phenomenology of the radion $\phi$, consider the case where it couples to 
gravity only minimally.  Its interaction with SM fields is given by\cite{Csaki:2000zn}
\be
\mathcal{L}_{\text{radion}}=\frac{\gamma\phi}{v}T_\mu^\mu,
\ee
where $v=246$ GeV is the vev of the SM Higgs field, and
$\gamma=v/(\sqrt6\Lambda)\simeq1.6\times10^{-3}$.  Thus, the coupling is quite weak.
After electroweak symmetry breaking, one finds that\cite{Bae:2000pk} 
\bea
T_\mu^\mu&=&\sum_fm_f\bar ff-2m_W^2W_\mu^+W^{-\mu}-m_Z^2Z_\mu Z^\mu\nonumber\\
&&+\left(2m_h^2h^2-\partial_\mu h\partial^\mu h\right)+\frac{\beta_{\text{QCD}}\left(
g_s\right)}{2g_s}G_{\mu\nu}^aG^{a,\mu\nu}\nonumber\\&&+\frac{\beta_{\text{QED}}\left(
e\right)}{2e}F_{\mu\nu}F^{\mu\nu},
\eea
where $\beta_{\text{QCD}}$ and $\beta_{\text{QED}}$ are the QCD and
QED beta functions, respectively. 

The dominant production channel for the 
radion at a hadron collider is via gluon-gluon fusion. For 
such a light radion, the strongest constraints are from exotic scalar searches  from  heavy meson decay processes, $M\to\gamma
\phi\to \gamma f\bar f$
where $M=\Upsilon(nS), J/\psi$. From the expressions in \cite{Nason:1986tr}
we find 
\be
\frac{\Gamma\left(M\to\gamma\phi\right)}{\Gamma\left(M\to e^+e^-\right)}
\lesssim10^{-8}
\ee
for $m_{\text{radion}}<4$ GeV and $M=\Upsilon(nS), J/\psi$. Such a 
tiny branching ratio is beyond current experimental sensitivity\cite{Love:2008aa,
Lees:2011wb,Lees:2012iw,Lees:2013vuj,Lees:2015jwa,Ablikim:2015voa}. 

\section{Standard Model matter in the bulk}
The SM particles might also propagate, or have non-trivial distributions,  
in some of the extra dimensions. For example, in the bulk RS model,
the coupling strengths between the KK graviton and the SM particles
are corrected by an overlap between the wavefunctions of the SM 
particles and the KK graviton in the extra dimensions \cite{Davoudiasl:2000wi}. 
Such corrections can be different for SM fermions ($C_{f\bar fG_1^*}$) and 
SM gauge bosons ($C_{VVG_1^*}$). If the correction increases 
$C_{VVG_1^*}/C_{f\bar fG_1^*}$, the gluon-gluon 
initial state contributes more to graviton production, and 
the production rate at 8 TeV LHC is decreased relative to the discussion above. 
Hence, constraints from 8 TeV LHC data are weakened.
Also, the $G_1^*$ decay branching ratio to the dilepton final state would 
decrease. Then, the dilepton signal might be much weaker than the 
diphoton signal, and might not be observed at 13 TeV LHC in the near future. 
There is a lot of model-dependency on the extra-dimensional geometry in such 
more-complicated models, which makes precise prediction here more difficult.

\section{Summary and conclusions}
We have found that a warped KK graviton, with a coupling strength 
given by a scale $\Lambda\simeq60^{+12}_{-10}$ TeV, is consistent with 
the 750 GeV excess observed at ATLAS and CMS.  Constraints from 
other channels, particularly the 13 TeV dilepton data, put pressure on 
such a result in the simplest warped compactification scenarios, but 
are not strongly inconsistent with it.  Thus, if the signal continues 
to be observed in future data, confirmation of such a scenario should 
also  readily  come from the dilepton channel.  The  constraints and this prediction are weakened for more complex warped scenarios.
But in either case, confirmation would also ultimately come  from angular 
distributions.  

In the event the current excess does not persist, the preceding 
discussion gives a simple parameterization of KK graviton phenomenology, based on a simplified model that derives from a general warped compactification scenario.  We have outlined how phenomenological parameters are related to the fundamental ones of such a warped compactification, and to aspects of the higher-dimensional geometry.  This discussion could thus be pertinent to unraveling such a signal in the event one is discovered at higher mass.

\section{Acknowledgments}
This work was supported in part by the U.S. DOE under Contract No. 
DE-SC0011702, and by Foundational Questions Institute (fqxi.org) 
grant FQXi-RFP-1507.  We thank N. Craig and L.-T. Wang 
for helpful discussions, and especially J. Incandela for a number of valuable suggestions and discussions, and for comments on a draft of this paper.

\appendix
\bibliographystyle{apsrev}
\bibliography{draft}
  
\end{document}